# Magnetic Properties and Large Coercivity of Mn$_x$Ga Nanostructures

M.E. Jamer[1], B.A. Assaf[1], S.P. Bennett[2], L.H. Lewis[3], and D. Heiman[1]

[1] Department of Physics, Northeastern University, Boston, MA 02115
[2] Department of Electrical Engineering, Northeastern University, Boston, MA 02115
[3] Department of Chemical Engineering, Northeastern University, Boston, MA 02115

**Abstract**

To investigate structure-property correlations, high-coercivity Mn$_x$Ga nanoparticles were synthesized by the method of sequential deposition of Ga and Mn fluxes using molecular beam epitaxy. Spontaneous nanostructuring was assisted by the use of an Au precursor and thermal annealing, and the growth properties, structure and magnetic properties were characterized. Atomic force microscopy revealed average particle dimensions of 100 nm and X-ray diffraction revealed a dominant tetragonal D0$_{22}$ crystal structures. Magnetic characterization at room temperature identified the presence of two magnetic phases, dominated by a high-coercivity (2.3 T) component in addition to a low-coercivity component.

## 1. Introduction

The binary Heusler compound Mn$_x$Ga (x~2-3) has recently gained interest due to attractive properties for magnetoelectronic device applications and as high-coercivity permanent magnets. The D0$_{22}$ phase of Mn$_x$Ga has been studied because of its half-metallic spin polarization, high Curie temperature and large magnetic anisotropy[1,2]. For Mn compositions of x = 2-3, the tetragonal D0$_{22}$ phase is ferrimagnetic, which has a low saturation magnetization that is suitable for spin transfer torque memory applications [3]. Magnetic studies of thin film Mn$_x$Ga with the D0$_{22}$ structure have revealed large coercivities [4,5,6,7] in excess of 2 T at room temperature attributed to a large magnetocrystalline anisotropy (K ~ 10 Merg/cm$^3$). Such high magnetocrystalline anisotropy makes Mn$_x$Ga a possible alternative to rare-earth and noble metal-based magnets in some applications [8], such as for heat-assisted magnetic recording (HAMR) media that will require nanostructured thin films exhibiting a high coercivity of approximately 3T.[9,10] Thin films of spontaneously nanostructured Mn$_x$Ga grown on Si substrates demonstrated a high coercivity of $\mu_o H_c$ = 2.5 T at 300 K. [6] This result provides us with an incentive for further investigation of the effects of nanostructuring on the magnetic properties of Mn$_x$Ga.

The present work focuses on synthesizing and characterizing the magnetic properties of Mn$_x$Ga nanostructures. The vapor-liquid-solid (VLS) growth method and sequential deposition of Mn and Ga by molecular beam epitaxy (MBE) were used for producing spontaneous nanostructures.

## 2. Preparation and Characterization of Nanoparticles

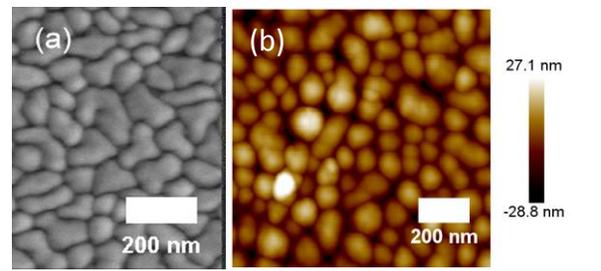

FIG. 1. (color online) (a) SEM image of $Mn_{2.5}Ga$ nanoparticles grown on Si.[6] (b) AFM image of $Mn_xGa$ grown with Au catalyst and sequential Ga and Mn deposition.

Realizing high coercivity requires structuring at the nanoscale in order for any inherent large magnetic anisotropy to be effective.[11] In this respect, single-magnetic-domain nanoparticles would have the highest coercivity.[6] Therefore, the focus of this study was to investigate the effect of structural modifications on the coercivity of $Mn_xGa$ by modifying the architecture of the nanoparticles. The VLS method requires a catalyst, normally Au, to form nanowires and nanoparticles by supersaturation and subsequent seed-mediated growth.[12] Here, the VLS method was not used to grow nanowires, but instead, Au was predeposited on the substrate in order to improve the uniformity of nucleation and growth of $Mn_xGa$ nanostructures.

It was found that the Au catalyst led to a more uniform distribution and shape of the resultant $Mn_xGa$ nanostructures. Samples were made by predepositing a 0.5 nanometer layer of Au on the Si substrate using electron-beam evaporation. The substrate was then heated to 550 $^oC$ at ~$10^{-9}$ Torr causing discrete Au nanoparticles to form. Elemental Ga and Mn were then deposited sequentially by MBE on the substrates held at 400 $^oC$. The flux ratio of Mn/Ga was set to give x = 2.5. The sequential deposition of Ga followed by introduction of Mn flux was used to overcome the low miscibility of Mn and Au at the growth temperatures. Although Mn does not readily enter the Au liquid without the presence of Ga [13], Mn can enter the Au-Ga liquid where it will form $Mn_xGa$.

The $Mn_xGa$ nanoparticles were studied by scanning electron microscopy (SEM), atomic force microscopy (AFM) and magnetic force microscopy (MFM). The composition of resulting nanoparticles was analyzed through SEM energy dispersive spectroscopy (EDS). The crystal structure of the nanoparticles was determined by X-ray diffraction (XRD) measurements. Magnetic characterization was performed using a Quantum Design superconducting quantum interface device (SQUID) magnetometer and a vibrating sample magnetometer (VSM) with a maxima of 5 T and 9 T applied field, respectively, at 300 K.

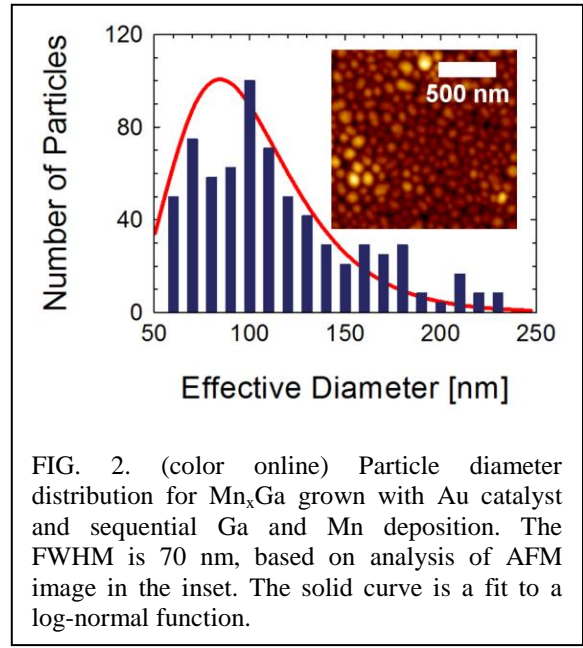

FIG. 2. (color online) Particle diameter distribution for $Mn_xGa$ grown with Au catalyst and sequential Ga and Mn deposition. The FWHM is 70 nm, based on analysis of AFM image in the inset. The solid curve is a fit to a log-normal function.

In previous work on nanostructured $Mn_{2.5}Ga$ thin films synthesized by MBE, the resulting particles did not have uniform shapes, as seen in the SEM image in Fig. 1(a) [6]. In contrast, the AFM image in Fig. 1(b) shows that the present growth of $Mn_{2.5}Ga$ assisted by Au nanoislands precursors results in particles that are more circular in cross section. These nanoparticles are quasispherical as shown in the AFM image. In this image the diameter of the particles are in the range 50-150 nm and the average particle height is approximately 50 nm. The average particle diameter and size distribution was determined by analyzing approximately 200 nanoparticles in the AFM image seen in the inset of Fig. 2. The areas of individual nanoparticles were measured and then converted into a distribution of mean diameters by $A = (\pi/4)D^2$ and then plotted in Fig. 2. A log-normal distribution [14] was used to determine the probability distribution given by

$$f(D) \sim (1/\sigma D) \exp\{-[\ln(D) - \mu]^2 / 2\sigma^2\}, \qquad \text{Eq. (1)}$$

where the arithmetic mean (expected value) for the particle diameter is $\langle D \rangle = \exp(\mu + \sigma^2/2)$ and standard deviation of the diameter distribution is $W = \exp(\sigma)$. The solid curve in Fig. 2 shows a plot of $f(D)$ with a mean diameter $\langle D \rangle = 110$ nm and full-width at half-maximum FWHM = 70 nm, determined from the fitting values $\mu = 4.68$ and $\sigma = 0.35$.

### 3. Crystal Structure Determination

Mn-Ga compounds exist in several crystal structures depending on the Mn:Ga ratio and temperature.[13] At low Mn concentration, $x < 1.6$, an $L1_0$ structure is preferred.[1,15]. For higher concentrations, $1.6 < x < 3$, the structure can be an antiferromagnetic *hcp* $D0_{19}$ phase, a cubic compensated ferrimagnetic $D0_3$ structure (disordered binary type of the $L2_1$)[16] when annealed above 600 °C, or a metastable tetragonal $D0_{22}$ ferrimagnetic phase for lower annealing temperatures.

The XRD pattern for a $Mn_xGa$ sample of this study is shown in Fig. 3(a). Strong $Mn_xGa$ Bragg peaks are indexed to the $D0_{22}$ crystal structure. Smaller Bragg peaks are observed for the $D0_3$ and $D0_{19}$ crystal structures. We note that the $D0_{22}$ peaks account for the majority of the integrated XRD intensity in this sample. Rietveld refinement analysis [17,18] showed that the $D0_{22}$ phase accounted for ~80% of the composition, with the $D0_{19}$ and $D0_3$ each accounting for ~10%.[16,19] The tetragonal $D0_{22}$ phase is found to possess lattice constants $a = 3.81 \pm 0.05$ Å and $c = 7.175 \pm 0.007$ Å, consistent with a Mn concentration of x=2.3-2.6.[2] Several minor XRD peaks are also observed, arising from cubic Au, *bcc* Mn and face-centered tetragonal Ga resulting from sequential deposition.

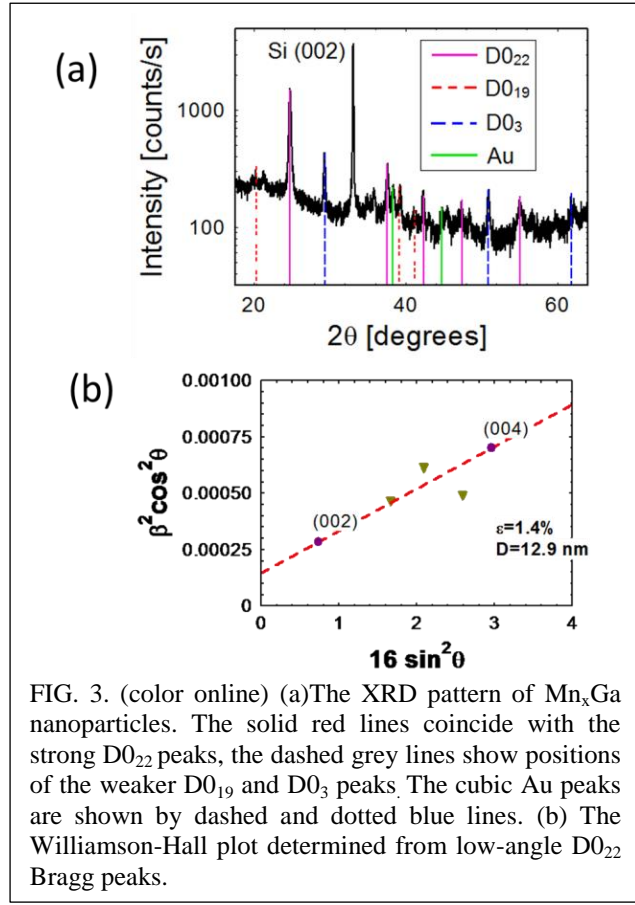

FIG. 3. (color online) (a)The XRD pattern of $Mn_xGa$ nanoparticles. The solid red lines coincide with the strong $D0_{22}$ peaks, the dashed grey lines show positions of the weaker $D0_{19}$ and $D0_3$ peaks. The cubic Au peaks are shown by dashed and dotted blue lines. (b) The Williamson-Hall plot determined from low-angle $D0_{22}$ Bragg peaks.

Chemical ordering of the D0$_{22}$ crystal sublattices may be estimated by examining the ratio of the Bragg peak intensities. The chemical order parameter $S$, which may vary from unity (fully chemically ordered) to zero (no chemical order) is defined through comparison of the experimental and computed XRD intensities using

$$S = \sqrt{[I^m_{(112)}/I^m_{(200)}]} / \sqrt{[I^c_{(112)}/I^c_{(200)}]}, \quad \text{Eq.(2)}$$

where $I^m$ are the experimental intensities and $I^c$ are the computed intensities. This procedure was applied to examine the sublattice disordering of the D0$_{22}$ structure.[20,21] For the (112) and (224) D0$_{22}$ Bragg peaks, the experimental intensity ratio $I^m_{(112)} / I^m_{(200)} = 1.27 \pm 0.2$ and the computed intensity ratio $I^c_{(112)} / I^c_{(200)} = 1.84$, provide a relatively high order parameter $S = 0.83 \pm 0.03$.

The Williamson-Hall analysis was used to determine the presence of strain in addition to any finite size effects from the diffraction data. The experimentally-observed broadening for Gaussian peaks is given by the sum of the particle size and strain contributions, respectively,

$$\beta^2 = (k\lambda)^2/(D\cos(\theta))^2 + 16\varepsilon^2 \tan^2(\theta), \quad \text{Eq. (3)}$$

where $\beta$ is the FWHM of the peak in radians, $\lambda=1.54$ Å is the x-ray wavelength, D is the coherent dimension of the particles and $\varepsilon$ is the strain.[22] In Eq.(3) the Scherrer constant is approximated as $k = 1$ and any instrumental broadening is neglected. In Figure 3(b), the Williamson-Hall plot of the product $\beta^2\cos^2\theta$ versus $16\sin^2(\theta)$ for the (002), (004), (112) and (200) Bragg peaks shows that the resulting particle size is D = 12.9 nm and the strain is $\varepsilon = 1.4$ %. It is assumed that there are possible atom dislocations in the D0$_{22}$ particles due to the relief of strain by atom dislocations.[22,23]

### 4. Magnetic Characterization

The magnetic properties of nanostructured Mn$_x$Ga samples were analyzed using SQUID and VSM magnetometry. Figure 4 shows the field dependence of the magnetic moment, m(H), illustrated by the solid (black) curve, where a wide hysteresis loop is observed with a coercive field of ~ 1 T at 300 K. Figure 4(upper inset) illustrates the high-field data up to 9 T, which indicates that the hysteresis loop saturates at 5 T. The field dependence of the sample at lower applied fields, where the hysteresis loop narrows, signals the presence of two magnetic phases, with low- and high-coercive components. We first note that there is a near-vertical rise in m(H) at very low fields indicating a

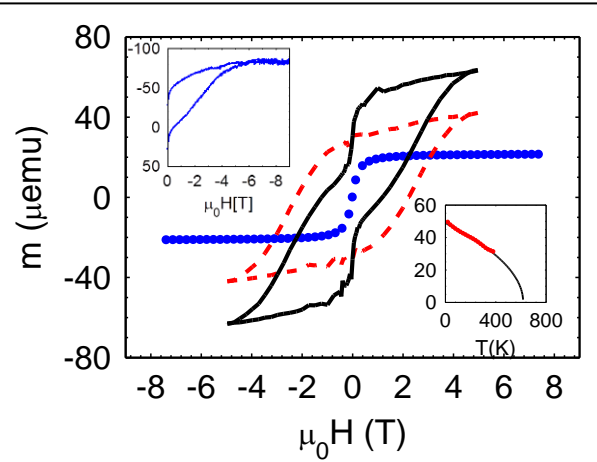

FIG. 4 (color online) Magnetic moment (black) as a function of applied magnetic field at room temperature for nanostructured Mn$_x$Ga grown with Au catalyst and sequential Ga and Mn deposition. The dashed (red) and dotted (blue) curves are low- and high-coercivity components, respectively. Inset (upper) shows m(H) to 9 T. Inset (lower) shows the low-field (0.1 T) moment versus temperature, where the extrapolation give T$_c$ ~ 600 K.

component that has negligible hysteresis ($H_c \sim 0$ T). This component can be modeled approximately by

$$m_L(H) = \frac{2m_L^0}{\pi}\arctan\left(\frac{\pi H}{2H_0}\right), \quad \text{Eq.(4)}$$

where $m_L^0$ is the remanence of the second component and $H_o$ is a scaling factor so that the moment saturates quickly.[24] The arctangent function is a simple antisymmetric function that reaches an asymptotic value smoothly with no discontinuities, which is similar to a noncoercive ferromagnetic M-H loop. The parameters $m_L^0 = 2.15 \times 10^{-5}$ emu and $H_0 = 0.19$ T have been chosen to fit the data and are shown by the dotted (blue) curve. The high-coercive component is obtained by subtracting the low-coercive component from the data. The resulting high-coercive component is shown by the dash (red) curve which has a coercive field of $\mu_o H_c = 2.3$ T, similar to the value found previously for $Mn_xGa$ nanostructures.[6]

The high-coercive component is attributed to the phase with the $D0_{22}$ structure in the sample. For single domain particles, the Stoner-Wohlfarth model predicts the maximum coercive field $H_c \sim 2K/M_s$.[25] The values for $D0_{22}$ $Mn_xGa$ (x = 2.5) for the anisotropy constant K are quite large and the values for $M_s$ are relatively small, which will lead to a large coercivity.[26] The experimental values for $Mn_{2.5}Ga$ at 300 K are $K = 12$ Merg/cm$^3$ and $M_s = 250$ emu/cm$^3$, which lead to a theoretical $H_c \sim 10$ T.[27] This calculation is the uppermost limit of the coercive field; however, the coercive field will decrease for multiple-domain nanoparticles or if strain is introduced to the system. The low-coercive component may be attributed to other phases apparent in the XRD pattern or the existence of strain caused by polycrystalline ordering of the $D0_{22}$ nanoparticles affecting the nanoparticles' anisotropy.

The magnetism of the sample is quite robust, as the Curie temperature is well above room temperature. The lower inset of Fig. 4 shows the low-field magnetic moment as a function of temperature, m(T). The lower solid curve is a fit of the data to $m(T) = m_o(1 - T/T_c)^{1/2}$, where the Curie temperature $T_c$ is approximately 600 K.

5. **Correlated Interaction Domains in Nanoparticles**

Figure 5 shows the AFM and related MFM images of the nanoparticles. In Fig. 5(a) the topographic image shows a fairly uniform area of nanoparticles that was previously subjected to a field of 9 T. The magnetic image in Fig. 5(b) shows the domain map of the nanoparticles, where light (positive phase) and dark (negative phase) areas correspond to opposite directions of the perpendicular component of the magnetic polarization. The positive and negative phase components correlate to one or multiple nanoparticles where the total diameter is D ~ 100 nm. When multiple nanoparticles correspond

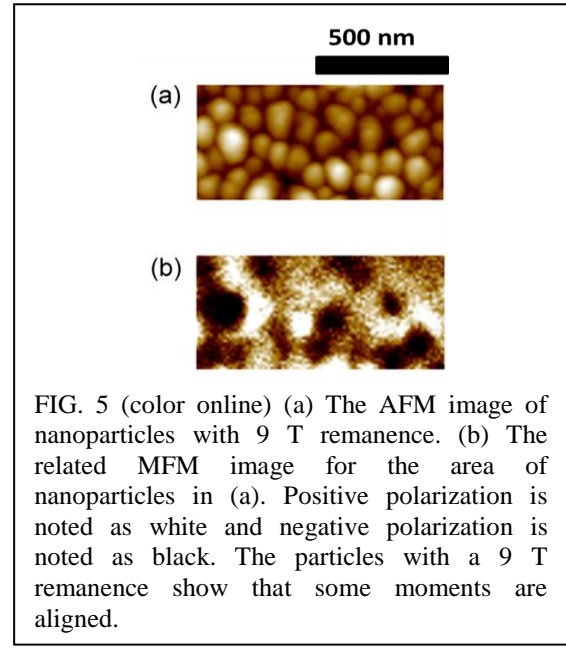

FIG. 5 (color online) (a) The AFM image of nanoparticles with 9 T remanence. (b) The related MFM image for the area of nanoparticles in (a). Positive polarization is noted as white and negative polarization is noted as black. The particles with a 9 T remanence show that some moments are aligned.

to one domain (both positive and negative) it is called interaction domains. The domains do not seem to be aligned with a preferred direction in Fig. 5(b). The nanoparticles (D ~ 50-100 nm) again show correlated domains consisting of 2-3 nanoparticles and show misalignment of domains. The MFM images in Fig. 5(b) indicate that the magnetic moments in different structures could be oriented in different directions based on the growth of the $Mn_xGa$ nanoparticles.

## 6. Conclusions

Nanostructured $Mn_xGa$ films grown with Au assistance for nucleation and sequential deposition of Mn and Ga produces spontaneous growth of $Mn_xGa$ nanostructures having a large coercivity, $H_c$, that are correlated into interaction domains. Several Mn-rich phases of $Mn_xGa$ (x=2-3) were observed, but the phase with the $D0_{22}$ structure was the dominant phase and underlies the 2.3 T coercive field measured at room temperature. The nanoparticles are approximately circular and fairly uniform in size, ~ 110 ± 35 nm diameter. The possible applications for these magnetic nanoparticles include rare-earth-free magnets and heat-assisted magnetic recording.

## Acknowledgements

We thank F. Jimenez-Villacorta and M. Loving for helpful discussions and A. Friedman for Au deposition. This work supported by a grant from the National Science Foundation DMR-0907007, and M.E.J. acknowledges support from the Northeastern ADVANCE program.